\documentstyle[11pt,newpasp,twoside]{article}
\def\edcomment#1{\iffalse\marginpar{\raggedright\sl#1\/}\else\relax\fi}
\reversemarginpar

\begin{document}
\title{Radiating Regions in Pulsar Magnetospheres: From Theory 
to Observations and Back}
 \author{Vladimir V. Usov}
\affil{Department of Condensed-Matter Physics, Weizmann Institute of
Science, Rehovot 76100, Israel}

\begin{abstract}
We discuss plausible locations of radiating regions in the magnetospheres 
of pulsars and argue that the mechanisms of pulsar radiation 
at different frequencies are closely connected with
the locations of the radiating regions, especially in the radio range.
If the region that is responsible for 
the bulk of the non-thermal radiation at some frequency
is localized in the pulsar magnetosphere then
the nature of this radiation will be either determined or, at least,
restricted significantly.

\end{abstract}

\section{Introduction}

Over 30 years since the discovery of pulsars, the mechanism of their 
radio emission is  still poorly understood. Moreover, the location
of radio emitting regions in the pulsar magnetospheres is unknown.
Strong non-thermal high-frequency (optical, X-ray 
and $\gamma$-ray) radiation is observed from about ten pulsars. 
The pulses of high-frequency radiation generally bear little 
resemblance to the radio pulses (with the exception of the Crab pulsar).
Besides, at different frequencies the pulses of the high-frequency radiation
are different. These imply
that there are many regions in the pulsar magnetospheres where
strong non-thermal radiation is generated. 

\section{Radiating Regions in Pulsar Magnetospheres}

A common point of all acceptable models of pulsars is that 
a strong electric field ${\bf E}_\parallel =({\bf E\cdot B})
{\bf B}/|{\bf B}|^2$ along the magnetic field ${\bf B}$
is generated in the pulsar magnetosphere.
Primary particles are accelerated by this electric field
to ultrarelativistic energies and generate $\gamma$-rays. 
Some of these $\gamma$-rays are absorbed by creating
secondary electron-positron pairs. The created pairs
screen the electric field ${\bf E}_\parallel$ in the pulsar
magnetosphere everywhere except for compact regions.
The compact regions where ${\bf E}_\parallel$ is unscreened are
called gaps. These gaps are "engines" that are
responsible for non-thermal radiation of pulsars. 
Most probably, the gaps are located either near the
magnetic poles of pulsars or near their light cylinders
(for a review, see Michel 1991). Some part of radiation
generated in the gaps and their vicinities escapes from 
the pulsar magnetospheres (see below) and may be observed.
Plasma instabilities may be developed in the
secondary electron-positron plasma. The regions of their
development are also
plausible sources of the non-thermal radiation of pulsars
in addition to the gaps.

\subsection{Radiation from Polar Gaps}

Gaps that form near the magnetic poles of pulsars
are called polar gaps. Physical processes (acceleration of 
particles, generation of radiation and its propagation, creation of
electron-positron pairs, etc.) in the polar gaps and their vicinities 
have been discussed in (e.g., Usov 1996; Zhang \& Harding 1999). 
The bulk of the radiation from the polar gaps is in
the range of hard X-rays and $\gamma$-rays. 
In all conventional polar-gap models (Ruderman \&
Sutherland 1975; Arons 1981; Zhang \& Harding 1999 and references
therein) where created electron-positron pairs are free 
the total power carried away by both relativistic particles and
radiation from the polar gap into the pulsar magnetosphere is the 
same within a factor of 2-3 or so. This power is about
ten times less than the $\gamma$-ray luminosity, at least for
the most part of pulsars detected in $\gamma$-rays.

It was shown (Shabad \& Usov 1986) that in a strong magnetic field,
$B > 0.1 B_{\rm cr}$, $\gamma$-rays emitted nearly along curved 
magnetic field lines adiabatically convert into bound pairs (positronium atoms)
rather than decaying into free pairs, where $B_{\rm cr}=4.4\times 10^{13}$ G.
This partially prevents the screening of the vacuum 
electric field near the pulsar surface,
and the total power carried away by both relativistic particles and
radiation from the polar gap into the pulsar magnetosphere can
increase significantly. For pulsars with a very strong magnetic field at
their surface,  $B_{_{\rm S}} > 0.1 B_{\rm cr}$,
a modified polar gap model was developed (Usov \& Melrose 1995, 1996).
In this model, the non-thermal luminosity of pulsars may be comparable
to the rate of rotational energy losses that is enough to
explain the observed luminosities of all known $\gamma$-ray pulsars.

A maser version of linear acceleration emission was suggested as a
mechanism of radio radiation from the polar gaps of pulsars 
(e.g., Melrose 1978; Rowe 1992). This mechanism requires an oscillating 
electric field ${\bf E}_\parallel$
in the polar gaps. The characteristic frequency of the radio
emission is $\sim \omega_0\Gamma^2$, where $\omega_0$ is the
oscillation frequency, and $\Gamma$ is the Lorentz-factor of radiating 
particles. In this model, the radio beam coincides 
with the beam of high-frequency (X-ray and $\gamma$-ray) emission from 
the polar gaps.

\subsection{Plasma Instabilities in Pulsar Magnetospheres
and Non-thermal Radiation from the Regions of their Development}

{\it Two-stream instability}. It was argued that the process of
pair creation is strongly nonstationary, and 
the pair plasma that flows away from the pulsar surface 
is nonhomogeneous and gathers into separate closes (e.g., Usov 1987).
Since the Lorentz-factors of the pair plasma lie within a wide range
(from $\Gamma_{\rm min}\sim 10$ to $\Gamma_{\rm max}\sim 10^4-10^5$),
the high-energy particles ($\Gamma\sim \Gamma_{\rm max}$) go ahead, and
the plasma clouds disperse as they go further from the pulsar.
At a distance of $\sim 2l\Gamma_{\rm min}^2$ from the pulsar surface
the high-energy particles of a given 
cloud catch up with the low-energy particles ($\Gamma\sim \Gamma_{\rm min}$)
that belong to the cloud going ahead of it, where $l$ is the
characteristic length between the clouds at the moment of their creation near
the pulsar surface. 
In the cloud overlapping region the energy distribution of particles
is two-humped, i.e., there are particles only with both $\Gamma\sim \Gamma_{\rm
min}$ and $\Gamma\sim \Gamma_{\rm max}$ whereas particles with intermediate
Lorentz-factors are absent. The plasma with such a distribution is 
unstable with respect to two-stream instability (Usov 1987; Ursov \&
Usov 1988; Asseo \& Melikidze 1998). Longitudinal (nonescaping) Langmuir 
waves that are generated in the process of development of this instability
may be converted by means of different non-linear effects
into electromagnetic waves that can escape from the
pulsar magnetosphere (e.g., Lesch, Gil, \& Shukla 1994;
Lyubarskii 1996; Mahajan, Machabeli, \& Rogova 1997; 
Melikidze, Gil, \& Pataraya 2000). This non-linear conversion of waves
is a "bottle-neck" that impedes the meeting of the model of pulsar radio 
emission and observational data.
The two-stream instability of strongly nonhomogeneous plasma develops
very fast, and if the process of
pair creation near the pulsar surface is strongly nonstationary indeed, 
the development of this instability in the pulsar magnetospheres 
is almost inevitable. It is worth noting that
the outflowing plasma may have several characteristic lengths of
its modulation. For instance, in the Ruderman-Sutherland model
they are $l_1\simeq 0.3 R$ and $l_2\simeq 2R_c$,
where $R$ is the neutron star radius and $R_c$ is the curvature radius
of magnetic field lines at the pulsar surface. For $R_c\simeq R\simeq 10^6$
cm and $\Gamma_{\rm min}\simeq 10$, the distances to the radio emitting
regions are $\sim 30R$ and $\sim 200R$. These distances are consistent
with observational data on pulsars (Rankin 1993; Kijak \& Gil 1997 and
references therein). In this model, 
high-frequency radiation from the radio emitting regions is very weak.

{\it Cyclotron instability}. Near the light cylinders of pulsars,
$r\sim c/\Omega$, the outflowing plasma may be unstable with respect to 
excitation of cyclotron waves (Machabeli \& Usov 1979), where $c$ is the speed
of light, $\Omega=2\pi /P$ and $P$ is the pulsar period. 
For typical pulsars, $B_{_{\rm S}}\sim 10^{12}$ G and $P\sim 1$ s, the frequency 
of these waves is in the radio range, from $\sim 10^2$ MHz to a few $\times
10^3$ MHz (Machabeli \& Usov 1989). The model that is based on the cyclotron
instability also can explain many observational data on radio emission of
pulsars (Lyutikov, Blandford, \& Machabeli 2000). The interaction between
cyclotron waves and outflowing particles leads to diffusion of particles
in the momentum space across the magnetic field. As a result, outflowing
particles acquire non-zero pitch angles and generate high-frequency
radiation via the synchrotron mechanism. The high-frequency luminosity of
the region where the cyclotron instability develops may be as high
as the power carried away by particles from the polar gap into the pulsar
magnetosphere.

\subsection{Radiation from Outer Gaps}

Charge deficient regions (outer gaps) with a strong electric field
$E_\parallel$ may exist near the pulsar light cylinder (e.g., Michel 1991).
The outer gap model describes the high-requency radiation of the Crab and
Vela pulsars fairly well (e.g., Cheng, Ho, \& Ruderman 1986a,b;
Romani 1996). Outer gaps may act as a generator of radiation 
in the pulsar magnetosphere only if the period of the pulsar rotation 
is small enough, $P<P_{\rm cr}\simeq$ a few $\times \,0.1$ s. 
Optical observations at the positions of $\gamma$-ray 
pulsars that are near the death line, $P\simeq P_{\rm cr}$, may test the 
outer gap model (e.g., Usov 1994; Lundqvist et al. 1999).

\section{Discussion}

For fast rotating pulsars, $P<P_{\rm cr}$, both polar and outer 
gaps can act in the pulsar magnetospheres, and therefore, the pair plasma
properties are
very uncertain. For typical pulsars with $P>P_{\rm cr}$, the polar gap model
has no an alternative. If for such pulsars it is confirmed that the distance $r$
from the pulsar to the radio emitting region is in the range 
$R\ll r \ll c/\Omega$ (Rankin 1993; Kijak \& Gil 1997 and references therein)
then most probably, the process of pair creation at the pulsar surface is
strongly nonstationary, and the two-stream instability of the outflowing
strongly nonhomogeneous plasma is a reason of the pulsar radio emission
(Usov 1987; Ursov \& Usov 1988; Asseo \& Melikidze 1998). 
However, if $r$ is either $\sim R$ or $\sim c/\Omega$, 
the most plausible mechanism of the pulsar radio emission is either

the linear acceleration emission (Melrose 1972) or the cyclotron 
instability (Machabeli \& Usov 1979, 1989; Lyutikov et al. 2000).

\acknowledgments

This research was supported by MINERVA Foundation, Munich / Germany.

\end{document}